\documentclass[aps,pra,twocolumn,nofootinbib,superscriptaddress,showpacs]{revtex4}
\usepackage{amsmath}
\usepackage{amsfonts}
\usepackage{amssymb}
\usepackage{graphicx}
\usepackage{comment}
\usepackage{color}

\begin{document}

\title{Finite key size analysis of two-way quantum cryptography}

\author{Jesni Shamsul Shaari}
\affiliation{Faculty of Science, International Islamic University Malaysia (IIUM),
Jalan Sultan Ahmad Shah, Bandar Indera Mahkota, 25200 Kuantan, Pahang, Malaysia}
\author{Stefano Mancini}
\affiliation{School of Science \& Technology, University of Camerino, I-62032 Camerino, Italy}

\date{\today}

\begin{abstract}
Quantum cryptographic protocols solve the longstanding problem of distributing a shared secret string to two distant users by typically making use of one-way quantum channel. However, alternative protocols exploiting two-way quantum channel have been proposed for the same goal and with potential advantages. Here we overview a security proof for two-way quantum key distribution protocols, against the most general eavesdropping attack, that utilize an entropic uncertainty relation. Then, by resorting to the `smooth' version of involved entropies, we extend such a proof to the case of finite key size. The results will be compared to those available for one-way protocols showing some advantages.
\end{abstract}


\maketitle
\section{Introduction}
\noindent Two-way quantum key distribution (QKD) schemes have evolved from a theoretic framework in the context of deterministic schemes to experimental realizations for QKD purposes \cite{TCS}. The protocol is described by a qubit being sent from one party, say, Bob to another, Alice for her encoding before being sent back to him for a measurement to ascertain Alice's encoding.
Standing in stark contrast to BB84 like prepare and measure schemes where information encoding is simply in the choice of states and bases of qubits to be sent from the encoding party to the receiver \cite{Gisin}, the case for two-way QKDs is really in the choice of unitary transformations by an encoding party that would act on the qubits traveling to and fro between the legitimate parties. This naturally makes use of the quantum channel twice.  Another peculiarity of the above-mentioned two-way protocols would be how its security is to be ensured. The protocols' runs are divided into two modes, namely the encoding mode (EM) where message encoded by Alice is followed by measurements by Bob for decoding purposes and the control mode (CM) where Alice would make projective measurements in randomly selected bases such that when the bases coincide with Bob's then errors can be ascertained. The distribution of the EM and CM would be determined by a factor $c$, with $0<c<1$, i.e. the probability of Alice randomly choosing an EM. 

However, despite its early introduction in 2002 \cite{PP}, and varying versions that followed, it was only about a decade later that a proper proof for one specific two-way protocol was found for unconditional security in \cite{c2w} and a proof for a purified version was reported in \cite{norm}. In \cite{c2w}, the proof was done for the specific case of non-entangled qubits in one of four states coming from two mutually unbiased bases (this protocol has been referred to as LM05 \cite{LM} and we shall refer to it as such hereafter). In \cite{norm}, a purification of two two-way QKD protocol, namely the Super Dense Coding scheme (SDC) and LM05 was done in  order to provide a security proof.

Unconditional security proofs provide for an information theoretic security picture of QKD. However, a very important fact that sets constraints within the operational context would be the issue of how such proofs are based on asymptotic analysis that holds true only for infinitely long keys; the concern regarding the security nature of realistic keys, which are finitely long, becomes evident. This has spurred a number of studies including \cite{haya,scar,tom}.

The central feature of this work is a review of the security analysis done based on protocol purifications in \cite{norm} and a natural generalization to the the case for finite sized keys. We begin with a quick description of two most relevant two-way QKD protocols followed by their purifications. After a brief on smooth entropies and its application to deriving finite key rates, we provide finite secure key rates in terms of efficiency for both protocols. These will then be compared to an asymmetrical BB84.


\section{Two-Way Protocols}

\noindent The SDC protocol in some sense is closer to the earlier instances of two-way protocols, namely the Ping-Pong \cite{PP} and more specifically its improved version \cite{qing} which provides for a higher protocol capacity in terms of the number of classical bits encoded in each EM run. It is however more resource demanding as it requires the use of a quantum memory on Bob's side. Our description of SDC here follows closely that of \cite{norm}. 
Very simply, the SDC protocol sees Bob sending half a Bell pair to Alice (while storing the other half in a quantum memory) for her to encode by virtue of a randomly chosen unitary transformation from the set containing the identity operator and the three Pauli operators, $\{I,\sigma_x,\sigma_y,\sigma_z\}$. She would subsequently submit the qubit back to Bob for measurements. Alternatively, she could  measure the received qubit in the $Z$ basis and prepare another in the $X$ basis to be sent to Bob. The former process, done with probability $c \approx 1$, corresponds to her actions in EM, while the latter, done with probability $1-c$, corresponds to CM.   Despite the use of $4$ unitaries imply a  larger alphabet used, particularly elements of $\mathbb{Z}_4$ $0,1,2,3$ mapped to $I,\sigma_x,\sigma_y,\sigma_z$ respectively  as opposed to bits for encoding, Alice could in fact assign logical bits (in pairs) $00,10,11$ and $01$ to the unitaries. However the mapping $f:\mathbb{Z}_4\rightarrow \mathbb{Z}_2^2$ to bits should be done only at the end of the protocol to avoid any possibility of Eve capitalizing on the correlation of bits given any pair \cite{bech}.

Bob on the other hand would, with a probability $c$ make a Bell measurement and with probability $1-c$ measure his stored and received qubit in the $Z$ and $X$ basis respectively. A protocol run which sees Bell measurements by Bob coinciding with Alice's unitary transformations would allow Bob to distinguish between Alice's unitaries perfectly (given that the Pauli matrices would shift between orthogonal Bell states). Assigning logical symbols $0, 1, 2$ and $3$ to the Bell states $|\psi^+\rangle,|\psi^-\rangle,|\phi^+\rangle$ and $|\phi^-\rangle$ respectively, these instances referred to as EM provide for sharing of a raw key between Alice and Bob. Again, it is possible for Bob to use bits instead when Alice submits the mapping information to Bob at the end of the protocol runs.  The instances when Alice's and Bob's measurements  in the $Z$ and $X$ bases coincides would allow for a meaningful CM. Hence a successful EM happens only with probability $c^2$ and CM happens with probability $(1-c)^2$. 

The case for a two-way QKD without using entanglement, namely the LM05, sees a qubit prepared by Bob in a particular bases randomly chosen from either the $X$ or $Z$ with probabilities $p_X$ and $p_Z$ respectively to be sent to Alice for her encoding. In the original LM05 protocol, apart from $p_X=p_Z$, only two unitary transformation was considered, namely the passive $I$ and $i\sigma_y$. In \cite{norm}, a modification was made to include another two unitary
transformation, namely the $\sigma_x$ and $\sigma_z$.  
The protocol was further presented in two versions, depending on the probabilities $p_X$ and $p_Z$.
In this protocol, upon executing her unitary transformation, Alice would resubmit the qubit to Bob who would measure in the same basis he prepared in. Defining the eigenstates of $\sigma_z$ and $\sigma_x$ as $|z\pm\rangle$ and $|x\pm\rangle$ respectively, bit values $0$ may be assigned to the states $|z+\rangle$ or $|x+\rangle$ and $1$ to the states $|z-\rangle$ or $|x-\rangle$. Bob then adds (modulo 2) the bit value corresponding to his prepared state (prior sending to Alice) to the bit value corresponding to his measurement result of the state (after Alice's encoding). We will mention shortly what Alice needs to do in order to share bit-wise information with Bob. This run of the protocol corresponds to a successful EM.

With the probability $1-c$, Alice would make measurements of the qubit she receives instead in either the $Z$ or $X$ bases for CM purposes.  

Apart from the use of non entangled qubits by Bob, another obvious difference with the SDC is in the possible options in the post processing feature of the protocols. In SDC, post processing only includes Alice's and Bob's discussion to ascertain which signals were to be used for EM and CM purposes apart from error estimations, error corrections and privacy amplifications. In the LM05 on the other hand, Alice and Bob have the option to commit to either direct reconciliation (DR) or reverse reconciliation (RR) which would determine the type of information to be disclosed over an authenticated public channel. In the case of RR, Bob should disclose the bases he prepared and measured qubits in.  Using the same assignment of logical bit pairs to unitary transformations as in SDC, Alice would keep only the first bit of the pair when Bob uses $Z$ and the second when Bob uses $X$. On the other hand, if a RR is considered, then Alice would reveal from which of the two sets, $S_0=\{I,i\sigma_y\}$ and $S_1=\{\sigma_x,\sigma_z\}$ was the transformation she had used and only the first bit of the bit pair for each transformation is used. This revelation is necessary given Bob's measurement in the $X$ basis would give him an erroneous bit when Alice uses $S_1$. In these cases, Bob would need to flip his bit.


\subsection{Purifications of Two-way Protocols}

\noindent The approach to the security proof presented in \cite{norm} is based on the notion of purification of protocols where the protocols can be described by the measurements made by Alice and Bob on a system provided to them by Eve \cite{LoChau}. Writing the measurements of Alice and Bob as Probability Operator Value Measure (POVM) maps $M_A^{\text{EM}}$ and $M_B^{\text{EM}}$ respectively in EM (or $M_A^{\text{CM}}$ and $M_B^{\text{CM}}$ in CM), the state after the measurements is given by 
\begin{eqnarray}
M_A^i\otimes M_B^i(\rho_{ABE}) 
\end{eqnarray}
with $i=\text{EM}$ or $\text{EM}$. In the purified version of both the SDC and the LM05 as reported in \cite{norm}, $\rho_{ABE}$ is a state in the Hilbert space $\mathcal{H}_A\otimes \mathcal{H}_B\otimes \mathcal{H}_E$ (Alice's, Bob's and Eve's respectively) and $\mathcal{H}_A$ and $\mathcal{H}_B$ are Hilbert spaces for two qubit systems each. 
Alice's measurement $M_A^{\text{EM}}$, in EM which acts on a two qubit system must be in such a way that it is equivalent to her encoding operation in the protocols. If we recall how in both protocols, Alice receives a state from Bob and resends after a unitary encoding on the state, $M_A^{\text{EM}}$ can be understood as a (POVM) acting on the received state with half of some entangled state so as the other half of the entangled state (after the measurement $M_A^{\text{EM}}$) is the same as the output from AliceÕs encoding to be sent to Bob\footnote{This is referred to as a purification of Alice's encoding in \cite{norm}.}. Thus we can imagine $\rho_{ABE}$ as that pure state which distributes a pair of qubits to Alice and Bob each. Alice's measurement on the two received qubits, $\text{tr}_{BE}(\rho_{ABE})$ would ensure that Bob's measurement, $M_B^{\text{EM}}$ on his received pair, $\text{tr}_{AE}(\rho_{ABE})$ is equivalent to Alice making a unitary transformation on Bob's prepared qubit for his subsequent decoding measurement.
With regards to Bob's measurements, $M_B^{\text{EM}}$, it is instructive to note that in SDC, $M_B^{\text{EM}}$ is really the Bell measurement for decoding purposes while in LM05, Bob's measurement on the first half of $\text{tr}_{AE}(\rho_{ABE})$ effectively prepares the qubit state sent to Alice (in the forward path) while his other measurement is on the other half of $\text{tr}_{AE}(\rho_{ABE})$.  The measurements $M_A^{\text{CM}}$ and $M_B^{\text{CM}}$ would then correspond to relevant local measurements of each qubit in the qubit pairs in CM (detailed in the ensuing section). Given these, the main ingredient in the security proof for the purified protocols is based on bounding Eve's information gain given Bob-Alice's using an uncertainty relation which measures the overlap of Bob's (Alice's) measurement apparatus in a reverse (direct) reconciliation scenario.


\subsection{Measurements and Entropic Uncertainty Relations}

\noindent Entropic uncertainty relations are used in some security proofs of particular QKD protocols given its power to describe bounds of uncertainty parties may share of a certain quantum system, say $B$. In its simplest description, it is simply Heisenberg's uncertainty principle in an entropic form first proven by \cite{hans}. It was later generalized \cite{EU1,EU2,EU3} to include correlation of the system to be measured with disjoint (possibly quantum) systems $A$ and $E$ given by  
\begin{eqnarray}\label{E1}
H(X|E)+H(Z|A)\geq -\log_2{\mathcal{C}},
\end{eqnarray}
where $X$ and $Z$ are results from POVMs $\mathbb{X}$ and $\mathbb{Z}$ respectively on $B$. Writing the POVM elements of $\mathbb{X}$ and $\mathbb{Z}$ as $\{E_x^\mathbb{X}\}$ and $\{E_z^\mathbb{Z}\}$ respectively, the term $\cal C$ is given by 
 \begin{eqnarray}\label{overlap}
\mathcal{C}= \max_{x,z}{\left\|\sqrt{E_x^\mathbb{X}}\sqrt{E_z^\mathbb{Z}}\right\|^2_{\infty}}.
 \end{eqnarray}
Eq.\eqref{E1} together with the Devatak-Winter security bound \cite{DW} is employed in \cite{norm} to provide a security proof for the purified SDC and LM05 protocols. In order to make use of such an entropic relation, \cite{norm} necessarily describes the term $\mathcal{C}$ as the (effective) overlap between Bob's measuring apparatus in a RR picture. 
In the purified SDC protocol, Bob's measurements are either the Bell measurements or a measurement that can be described as $\sigma_z\otimes \sigma_x$ where the overlap between the POVM elements as defined in eq.(\ref{overlap}) is $1/4$. This maximal overlap is achieved when considering the overlap between the POVM elements in EM and that of in CM. If we let the measurements by Alice and Bob to result in the strings $S_A^\text{EM}$ and $S_B^\text{EM}$ respectively from EM and $S_A^\text{CM}$ and $S_B^\text{CM}$ respectively from CM, and Eve's system as $E$ we can bound Eve's information based on eq.(\ref{E1}) as 
\begin{eqnarray}
H(S_B^\text{EM}|E)+H(S_B^\text{CM}|S_A^\text{CM})\geq 2.
\end{eqnarray} 
With $H(S_B^\text{CM}|S_A^\text{CM})$ upper bounded by $h_4(q_\text{CM})$ and $H(S_B^\text{EM}|S_A^\text{EM})$ by $h_4(q_\text{EM})$ where 
 $h_4$ is the 4-ary Shannon entropy,
$\overline q_\text{CM}$ and $\overline q_\text{EM}$ are errors in Alice's and Bob's strings in CM and EM respectively,
the key rate of the SDC, $R_{SDC}$, is given by the Devatak-Winter rate \cite{DW} 
\begin{eqnarray}
R_{SDC}\geq 2-h_4(\overline q_\text{CM})-h_4(\overline q_\text{EM}).
\end{eqnarray} 
It is instructive to note that the errors $\overline q_\text{CM}$ and $\overline q_\text{EM}$  affects a two-bit message transmission resulting from a use of two noisy channels and thus comes as a triple; namely error exclusively in either one channel, say $q^a_1$ and $q^a_2$ as well as errors in both channels, $q^a_{3}$ for $a=\text{CM},\text{EM}$. Hence, 
\begin{eqnarray}
h_4(\overline q_a)=-\sum_i^3q_i^a\log_2{q_i^a}-(1-\sum_i^3q_i^a)\log_2{\left(1-\sum_i^3q_i^a\right )}.
\end{eqnarray} 
It is certainly possible to consider the entropic bounds dictated by the overlap of Alice's apparatus instead of Bob's as shown above. The key rate would remain the same nonetheless. 
\newline
\newline
A similar line of argument can be applied to the LM05 protocol; 
though the measurements made for decoding purposes by Bob would be $\mathcal{Z}=\sigma_z\otimes \sigma_z$ or $\mathcal{X}=\sigma_x\otimes \sigma_x$ followed by an XOR of the the measurements' results to reveal the encoding done by Alice in the EM. Another set of measurements, which would be useful for the CM would be $\mathcal{Z}_C=\sigma_z\otimes I$ or $\mathcal{X}_C=\sigma_x\otimes I$ where the maximal overlap between the POVM elements for measurements made in EM and that in CM is $1/2$.
Using the relevant entropic bounds (similar to SDC though the overlap is now $1/2$ instead of $1/4$), the key rate for LM05, $R_{LM05}$ can be easily shown to be 
\begin{eqnarray}
R_{LM05}\geq 1-h_2(q_\text{CM})-h_2(q_\text{EM}),
\end{eqnarray} 
where $h_2$ is the binary Shannon entropy and $q_\text{CM}$ and $q_\text{EM}$ are the errors in the CM and EM respectively.

The above key rates are derived from inequalities based on terms that are meaningful only within the infinitely long key limits and is thus unsatisfactory from an operational and practical perspective.  In the following subsection, we shall review briefly the notion of smooth entropies and relevant entropic bounds that we shall use to derive a finite key rate for SDC and LM05.


\subsection{Smooth Entropies and Finite Keys}

\noindent The smooth entropy is defined based on the conditional entropy. More rigorously, following the definition given in \cite{toma} for a bipartite state $\rho_{BE}$ on $B$ and $E$, the entropy of $B$ given $E$ is defined as
\begin{eqnarray}
H_{min}(B|E)=\max_{\rho_E}{\sup{\{\lambda\in \mathbb{R}:2^{-\lambda} I_B\otimes \rho_E\geq \rho_{BE}\}}}
\end{eqnarray}
where the maximum is taken overall states $\rho_E$ in $E$ and $I_B$ is the identity on $B$. The $\epsilon$-smooth min-entropy for  $\epsilon\geq 0$ is then defined as 
\begin{eqnarray}
H^\epsilon_{min}(B|E)=\max_\rho{H_{min}(B|E)}.
\end{eqnarray}
with maximization over all bipartite states $\rho$ on $B$ and $E$ with a purified distance \cite{toma2} to $\rho_{BE}$ not exceeding $\epsilon$. The smooth max-entropy is defined as the dual of the smooth min-entropy with regards to any purification of $\rho_{BE}$.

For the tripartite state $\rho_{ABE}$ and POVMs $\mathbb{X}$ and $\mathbb{Z}$ respectively on $B$ (resulting in bit strings $X$ and $Z$), from \cite{toma}, the smooth min-entropy of $X$ conditioned on $E$, $H_{min}^\epsilon (X|E)$, gives the number of bits contained in $X$ that are $\epsilon$-close\footnote{According to a distance  that is based on the same notion of purified distance,  for the classical register $X$ arising from POVM $\mathbb{X}$ on $B$, given quantum side information $E$ \cite{toma3}} to a uniform distribution and independent of $E$. The smooth max-entropy of $Z$ conditioned on $A$, $H_{max}^\epsilon (Z|A)$, gives the number of bits needed to reconstruct $Z$ from $A$ up to a probability of failure $\epsilon$ and the generalized uncertainty relation involving smooth entropies is given as \cite{toma}
\begin{eqnarray}\label{E2}
H_{min}^\epsilon (X|E)+H_{max}^\epsilon (Z|A)\geq -\log_2{\mathcal{C}}.
\end{eqnarray}
In identifying the measurements and results  in both the SDC and LM05 with eq.(\ref{E2}), similar to the above, $H_{min}^\epsilon (X|E)$ will be identified with Eve's correlation with Bob's string (in EM)  while  $H_{max}^\epsilon (Z|A)$ is to be identified with Alice's and Bob's in CM. There are two points worth mentioning with regards to the measurements made in CM. The first is in particular reference to LM05's measurements  $\mathcal{Z}_C$ and $\mathcal{X}_C$ where a passive operation is noted on the second qubit (received in the backward path). This does not necessarily require Bob to \textit{not} measure the qubit in the backward path; rather he could just ignore the measurement made and consider only the result of his first measurement.\footnote{The results should coincide with Alice's measurement in CM when their bases coincide.}  In other words, in the cases Alice note a particular round of the protocol is a CM, Bob would ignore the result of his measurement on the second qubit. If it was the EM, then both the measurement results would be XOR-ed for decoding purposes. 

The second point is that as the $\sigma_z\otimes \sigma_x$ in SDC and $\mathcal{Z}_C$ and $\mathcal{X}_C$ in LM05 happens only in CM, one can see the protocol as analogous to the asymmetrical prepare and measure protocol of BB84 where the measurements in EM is seen as measurements in the preferred basis in the asymmetrical BB84. In the case for the latter, in \cite{tom}, where measurements are made in the $X$ and $Z$, a \textit{gedankenexperiment} was considered where all measurements were done in the $Z$ basis to establish an uncertainty relation. Following \cite{tom}, in the case for the LM05, we can use the bits derived in the CM to provide for an estimation of the errors in the application of the uncertainty relation using a similar \textit{gedankenexperiment} where all rounds are CM; and since Bob can choose to measure for EM, security follows from the notion that the better Alice could estimate Bob's bits in CM, the worse would Eve's estimation of Bob's bits in EM.


\section{Efficiency and Secure Key Rates}

\noindent In what follows, we shall assume implementations of the two-way protocols using depolarizing channels. The depolarizing channel, $\mathcal{D}$ is described by the parameter $q$, such that $0 \le q \le 1$, which affects a quantum state, $\rho$ independent of the basis as such 
\begin{eqnarray}
\rho\mapsto \mathcal{D}(\rho) = (1-q)\rho+qI/2.
\end{eqnarray}
For two-way protocols, the use of depolarizing channels can be categorized as either independent or correlated channels. Such a correlation can be understood in terms of the errors estimated in the forward and backward paths in CM against errors in EM. Given  bit wise errors $e_1$ and $e_2$ in the forward and backward paths respectively, we say the channels are independent provided the errors in EM is given by $e_m=e_1(1-e_2)+e_2(1-e_1)$. Otherwise they are correlated \footnote{A similar case for correlations between errors in the forward and backward path in CM was studied in \cite{SLM}}. While the case for independent channels are unique by definition, the cases for correlated channels can be infinitely many. However, we shall only consider, as in \cite{norm} correlated channels where $e_1=e_2=e_m$.   

To ascertain a finite key rate for the two-way protocols given such implementation, we need to determine the distribution of finite number of bits between the EM and the CM. Choices of the value of $c$ for both protocols would determine this and ultimately how much of a key rate one can have. Hence, in order to determine the optimal value for $c$, assuming one has a value of $k$ for bits in CM and $n$ in EM (more precisely the cases where Alice's encoding measurements coincide with Bob's decoding measurements),  the total number qubits, $M(n,k)$ sent\footnote{While for LM05 only a single qubit travels to and fro between the communicating parties, the case for SDC makes use of entangled pairs. Thus $M(n,k)$ for SDC must be understood as number of qubit pairs.} before $n$
bits are derived from EM and $k$ for CM estimation can be made is given by probability value $c=(\sqrt{1+k/n})^{-1}$. This is immediately derived from modelling the protocol based on \cite{tom}.

Subsequently, the already mentioned \textit{gedankenexperiment} will provide for an uncertainty relation of eq.(\ref{E2}) and the smooth-max entropy term for an $n$-bit string, $Z$ given $C$, $H^\epsilon_{max}(Z|C)$ would be upper bounded by $nh_2(Q+\mu(n,k))$ where  
\begin{eqnarray}
\mu(n,k)=\sqrt{\dfrac{n+k}{nk}\dfrac{k+1}{k}\ln{\dfrac{2}{\epsilon_S}}},
\end{eqnarray}
with the strings $Z$ and $C$ derived from CM differing at $Q$ bits. The parameter $\epsilon_S>0$ is a security parameter as defined in \cite{tom}. 

Strings shared between Alice and Bob in the EM, which one party's may be characterized as having errors $e$ relative to the other need be subjected to an error correction procedure. This can conventionally be understood as the amount of pre shared secret bits invested in the communications for error correction purposes\footnote{It is possible to consider a more practical scenario, for example for a bit string with error rate $e$, a cofactor would be multiplied to the amount bits needed for such a purpose given by $h_2(e)$.}. 

The Leftover Hashing Lemma \cite{hash} then provides for the length of the secret key as
\begin{eqnarray}\label{L}
L=\left \lfloor{H^\epsilon_{min}(X|E)+2\log{2\epsilon}}\right \rfloor,
\end{eqnarray}
which is $2\epsilon$ close to a bit string which is uniform and independent of Eve's knowledge. 

The efficiency of the protocol, $\mathcal{E}$ can be defined based on the amount of resources in terms of number of qubits required, 
\begin{eqnarray}
\mathcal{E}=L/M(n,k).
\end{eqnarray} 
It should be noted that for SDC, a factor of 2 should be further multiplied to the denominator reflecting the use of entangled pairs. As in \cite{tom}, we do not include any classical bits necessary in the protocol unlike \cite{cabello}.


\subsection{Finite Key Analysis for SDC}

\noindent We let a total of $2M(n,k)$ pairs of qubits travel to and fro between Bob and  Alice with the latter committing to encoding with probability $c$. Thus one gets $n$ amount of quaternary digits from the EM. Following earlier notation, the $n$ quaternary strings of Alice and Bob are $S_A^\text{EM}$ and $S_B^\text{EM}$ respectively while the $k$ quaternary strings derived from the CM are $S_A^\text{CM}$ and $S_B^\text{CM}$ respectively. 
Considering a \textit{gedankenexperiment} where all runs are CM lets us bound Eve's information on the $n$ quaternary string  using equation (\ref{E2}) as
\begin{eqnarray}\label{Hsdc}
H_{min}^\epsilon (S_B^\text{EM}|E)+H_{max}^\epsilon (S_B^\text{CM}|S_A^\text{CM})\geq 2n.
\end{eqnarray}
The term $H_{min}^\epsilon (S_B^\text{EM}|E)$ is the smooth min-entropy of $S_B^\text{EM}$ conditioned on $E$. It reflects Eve's correlation with Bob's string in the EM giving  the number of bits contained in $S_B^\text{EM}$ that are $\epsilon$-close to a uniform distribution and independent of $E$. In ascertaining the term $H_{max}^\epsilon (S_B^\text{CM}|S_A^\text{CM})$ reflecting the correlation between Bob's and Alice's strings, the case for two way channels with depolarization $q$ each allows us to consider bit error rate in qubit measurements as $q/2$ for each channel instead of a channel transmitting quaternary digits instead. Thus in this case,
\begin{eqnarray}
h_4(\overline{q})=2h_2(q/2),
\end{eqnarray}
for $\overline{q}=((1-q/2)q/2,(q/2)(1-q/2),q^2/4)$.
The number of bits required to reconstruct $S_B^\text{CM}$ from $S_A^\text{CM}$ (as two $n$ bit length strings) up to a probability of failure $\epsilon$,
\begin{eqnarray}\label{Hsdc2}
H_{max}^\epsilon (S_B^\text{CM}|S_A^\text{CM})\le 2nh_2\left(\dfrac{q}{2}+\mu(n,k)\right),
\end{eqnarray}
where $q/2$ is the error rate for each independent depolarizing channel in the CM.
If we write the smooth max-entropy of Alice's string conditioned on Bob's in EM as $H_{max}^\epsilon (S_B^\text{EM}|S_A^\text{EM})\le nh_4(\mathcal{Q}_E)$ with $\mathcal{Q}_E$ as errors in the EM, then equations (\ref{L}), (\ref{Hsdc}) and (\ref{Hsdc2}) gives 
the key length after error correction and privacy amplification as  
\begin{eqnarray}\label{KSDC}
L_{SDC}&\leq& n\left [2-2h_2\left(\dfrac{q}{2}+\mu(n,k)\right)-h_4(\mathcal{Q}_E)\right ]
-\log{\dfrac{2}{\epsilon_S^2}},\nonumber\\
\end{eqnarray}
where from \cite{tom}, we have the term $nh_4(\mathcal{Q}_E)$ as the amount of information leaked to Eve due to error correction of the string $S_A^\text{EM}$ and $\log{(\epsilon_S^2)}$ results from the quantum Leftover Hashing Lemma \cite{hash}.


\subsection{Finite Key Analysis for LM05}

\noindent In this work, we will only focus on the first version of the LM05 protocol where Bob's choice for qubit preparation and measurement in the $Z$ bases is preferred ($p_Z\approx 1$) and Alice either encodes with anyone of her transformations with probability $c/4$ or make measurements in the $X$ bases is with probability $1-c$. Furthermore, the deliberations ensue would be focused on the RR scenario.
Alice's and Bob's choice of measuring $2M(n,k)$ qubit pairs in the basis $X$ with probability $1-c$ results in $n_e=n+\sqrt{nk}$ number of bits derived from EM and $k$ for CM. Hence $\sqrt{nk}$ pairs would be wasted due to bases mismatch when Alice measures in $X$ while Bob chooses $Z$ (notice that $2\sqrt{nk}$ qubits are wasted in \cite{tom} for similar reasons). 

Let us write the string Bob has in the EM and CM as $s_B^\text{EM}$ and $s_B^\text{CM}$ respectively as well as Alice's as $s_A^\text{EM}$ and $s_A^\text{CM}$ respectively. 
As described above, let us consider a \textit{gedankenexperiment} where all qubits are measured using $X$ basis (for CM), we can therefore write an uncertainty relation from equation (\ref{E2}) to bound Eve's information as
\begin{eqnarray}\label{ELM05}
H_{min}^\epsilon (s_B^\text{EM}|E)+H_{max}^\epsilon (s_B^\text{CM}|s_A^\text{CM})\geq n_e.
\end{eqnarray}
Similar to earlier discussions, the first term in the left hand side of the above equation is the smooth min-entropy reflecting  the number of bits contained in Bob's string, $s_B^\text{EM}$ that are $\epsilon$-close to a uniform distribution and independent of Eve. The second term $H_{max}^\epsilon (s_B^\text{CM}|s_A^\text{CM})$ reflects the correlation between Bob's and Alice's strings and gives the number of bits needed to reconstruct $s_B^\text{CM}$ from $s_A^\text{CM}$ up to a probability of failure $\epsilon$,
\begin{eqnarray}\label{ELM052}
H_{max}^\epsilon (s_B^\text{CM}|s_A^\text{CM})\le n_eh_2\left (\dfrac{q}{2}+\mu(n_e,k)\right).
\end{eqnarray}
With $Q_f$ as the error rate in EM, noting $H_{max}^\epsilon (s_B^\text{EM}|s_A^\text{EM})\le n_eh_2(Q_f)$ and following equations (\ref{L}), (\ref{ELM05}), (\ref{ELM052}), the key length after error correction and privacy amplification is given by
\begin{eqnarray}\label{KLM05}
L_{LM05} & \leq & n_{e}-n_eh_2\left (\dfrac{q}{2}+\mu(n_e,k)\right)\nonumber\\
&&-n_eh_2(Q_f)-\log{\dfrac{2}{\epsilon_S^2}}.
\end{eqnarray}


\section{Numerical results and comparisons}

\noindent In making use of the key rate formulae of equations (\ref{KSDC}) and (\ref{KLM05}) above, we need to first assume a value for the security  parameter $\epsilon_S$, for which we set as $10^{-10}$. Then, for a given value for errors in the CM (errors in EM is then immediately defined), setting the number for $M(n,k)$, we determine the value $k$ which achieves the  maximal value for secure key length. As $M(n,k)$ approaches infinite for $n\rightarrow \infty$, following \cite{tom} we could let $k<\mathcal{K}\sqrt{n}$ for some fixed $\mathcal{K}$, so that
\begin{eqnarray}
\lim_{n\to\infty}{\mu(n,k)}=0,
\end{eqnarray}
and consider secure keys within the infinite key regime. In the infinite key regime, the efficiency for the protocols SDC becomes
\begin{eqnarray}
\lim_{n\to\infty}{L_{SDC}/2M(n,k)}=1-h_2\left(\dfrac{q}{2}\right)-\dfrac{h_4(\mathcal{Q}_E)}{2},
\end{eqnarray}
and for LM05 we have 
\begin{eqnarray}
\lim_{n\to\infty}{L_{LM05}/M(n,k)}=1-h_2\left (\dfrac{q}{2}\right)-h_2(Q_f).
\end{eqnarray}
It is worth noting that the `additional' term $\sqrt{nk}$ for amount of bits that can be derived from EM in LM05 becomes negligible in the infinite key regime and $\lim_{n\to\infty}{\sqrt{nk}/M(n,k)}=0$. Furthermore, for this condition, it would be only the final term (related to error correction) that would determine how one protocol outperforms another.


\subsection{Independent Channels}
\noindent In Figures 1 and 2, we present the results for implementations of the SDC and LM05  for independent channels compared to the asymmetrical BB84 in terms of protocol efficiency against errors. In this case, the error rates for SDC and LM05 in EM are given by $2q/2 (1-q/2)$ and $(2q-q^2)/4$ respectively \cite{norm}.
\begin{figure}[ht] 
   \centering
   \includegraphics[width=0.45\textwidth]{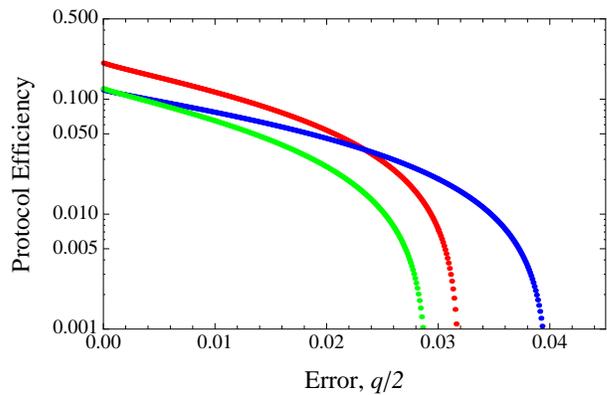} 
   \caption{Efficiency for Asymmetrical BB84 (blue), LM05 (red) and SDC (green) against error rate $q/2$ given independent channels for total qubit used as $M=10^4$.}  
   \label{fig1}
\end{figure}  

\begin{figure}[ht] 
   \centering
   \includegraphics[width=0.45\textwidth]{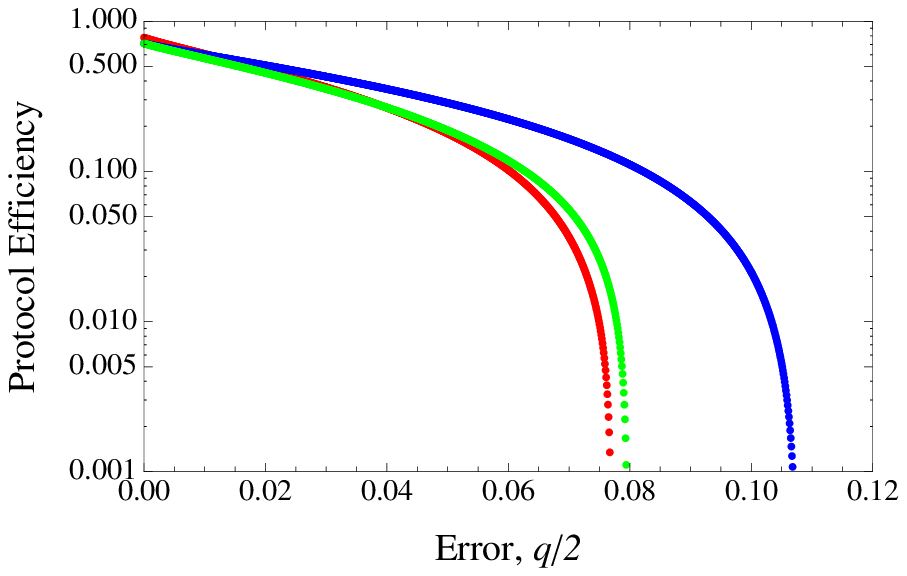} 
   \caption{Efficiency for Asymmetrical BB84 (blue), LM05 (red) and SDC (green) against error rate $q/2$ given independent channels for total qubit used as $M=10^7$.}
   \label{fig2}
\end{figure}   
\begin{figure}[ht] 
   \centering
   \includegraphics[width=0.45\textwidth]{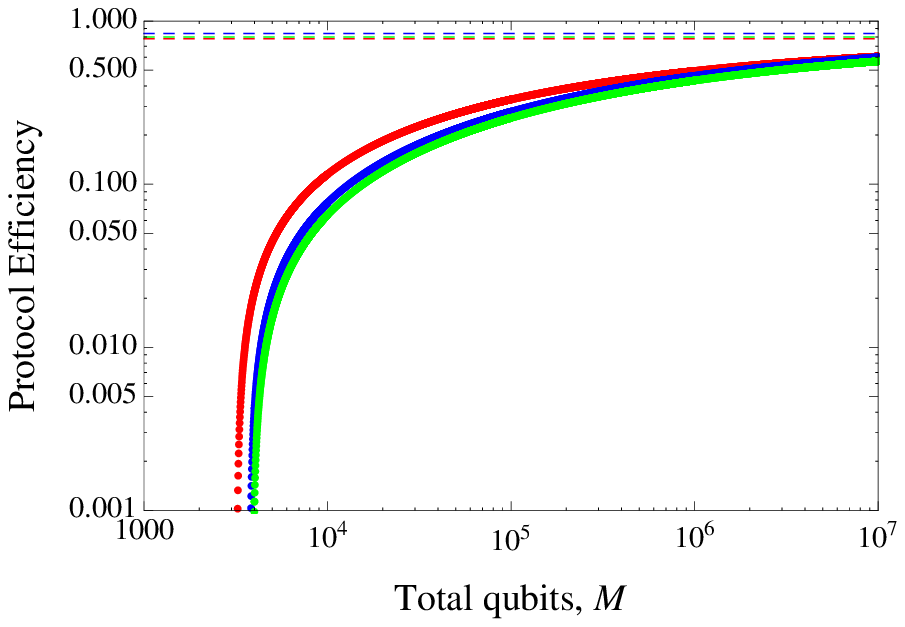} 
   \caption{Efficiency for Asymmetrical BB84 (blue), LM05 (red) and SDC (green) against total qubit used, $M$ (for SDC, it is to be understood as total pairs used) for error rate $0.01$. The horizontal dashed lines represents the infinite key regime.}  
   \label{figA}
\end{figure}  

\begin{figure}[ht] 
   \centering
   \includegraphics[width=0.45\textwidth]{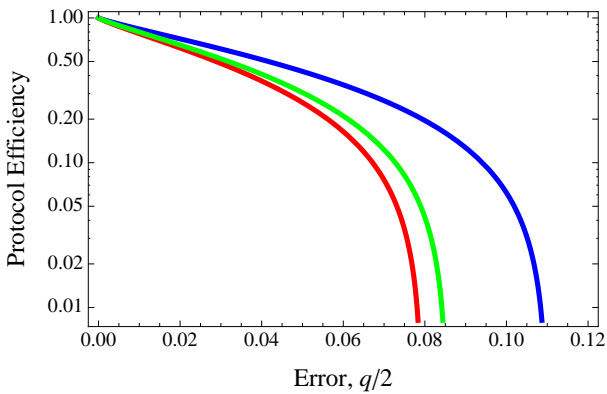} 
   \caption{Efficiency for Asymmetrical BB84 (blue), LM05 (red) and SDC (green) against error rate $q/2$ given independent channels  in the infinite key limit 
     }
   \label{fig4}
\end{figure}   

We can observe that there are particular regions where LM05 outperforms both SDC and BB84. Generally, this can be understood when considering the competing terms of the extra $\sqrt{nk}$ contributing to LM05's raw key versus the amount of bits to be discarded due to error correction. For LM05, this is given by $h_2[2q/2 (1-q/2)]$, SDC's $h_4[(2q-q^2)/4]/2$  while for BB84, it is simply $h_2(q/2)$ where 
\begin{eqnarray}
h_2[2q/2 (1-q/2)]\geq h_4[(2q-q^2)/4]/2\geq h_2(q/2),
\end{eqnarray}
with equality only at $q=0$ or $q=1$.
The term $\sqrt{nk}$ is most dominant for relatively small values of $n$, hence LM05's best performance of the three in Figure \ref{fig1} where $M(n,k)=10^4$. In the same figure we see instances when the errors are big enough and outweighs the contribution from  $\sqrt{nk}$, LM05 performs poorly compared to BB84. The error correction term for SDC is only very slightly better than that of LM05 in the infinite key regime, explaining why it still does not exceed LM05's $\sqrt{nk}$ advantage in the finite key scenario here. Thus LM05 exceeds BB84 up to an error rate of about $2.7\%$. However, in Figure \ref{fig2} where the number of qubits used, $M(n,k)=10^7$, LM05 exceeds BB84 only up to about $1\%$ and SDC's up to $3.8\%$. The plot of the protocols' efficiency against $M(n,k)$ for the error of $0.01$ in Figure \ref{figA}  exhibits the convergence of the efficiencies as the number of qubits used increases up to $10^7$. These results clearly emphasizes LM05's determinism over the asymmetric BB84's. It is worth recalling the fact that while LM05 claims deterministic status in terms of the absence for bases mismatch in EM, BB84 only approximates this (in the infinite key regime). Figure \ref{fig4} exhibits the case for the three protocols in the infinite key regime.


\subsection{Correlated Channels}
\begin{figure}[ht] 
   \centering
   \includegraphics[width=0.45\textwidth]{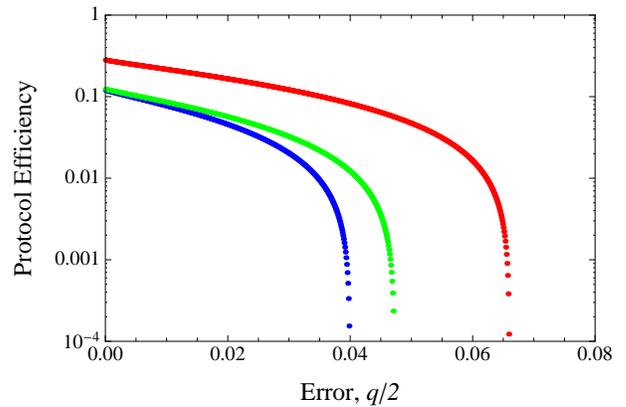} 
   \caption{Efficiency for Asymmetrical BB84 (blue), LM05 (red) and SDC (green) against error rate $q/2$ for total qubit used as $M=10^4$ for correlated channels specified by $e_1=e_2=e_m$.}
   \label{fig5}
\end{figure}

\begin{figure}[ht] 
   \centering
   \includegraphics[width=0.45\textwidth]{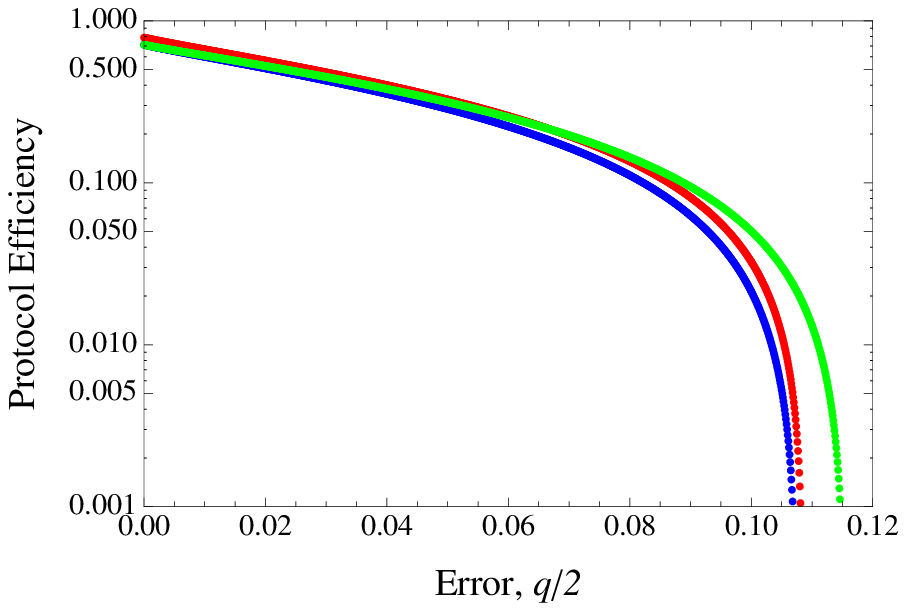} 
   \caption{Efficiency for Asymmetrical BB84 (blue), LM05 (red) and SDC (green) against error rate $q/2$ for total qubit used as $M=10^7$ for correlated channels specified by $e_1=e_2=e_m$.}
   \label{fig6}
   \end{figure}

\noindent In Figures \ref{fig5} and \ref{fig6} we present the results for implementations of the SDC and LM05  for correlated channels compared to the asymmetrical BB84 in terms of their efficiencies against errors in the channel. We reiterate that we consider a particular model for correlated channels; i.e. specified by $e_1=e_2=e_m$ and the error rates for SDC and LM05 in EM are given by $q/2$ and $q/4$ respectively \cite{norm}.
      
 \begin{figure}[ht] 
   \centering
   \includegraphics[width=0.45\textwidth]{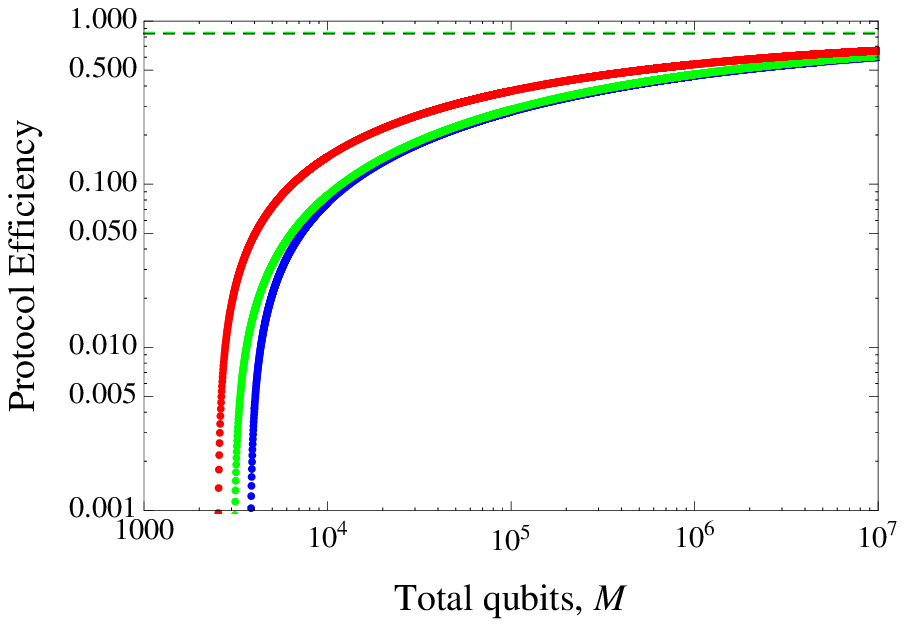} 
   \caption{Efficiency for Asymmetrical BB84 (blue), LM05 (red) and SDC (green) against total qubit used, $M$ (for SDC, it is to be understood as total pairs used) for error rate $0.01$ for correlated channels specified by $e_1=e_2=e_m$. The horizontal dashed lines represents the infinite key regime.}
   \label{fig7}
   \end{figure}   
   
   \begin{figure}[ht] 
   \centering
   \includegraphics[width=0.45\textwidth]{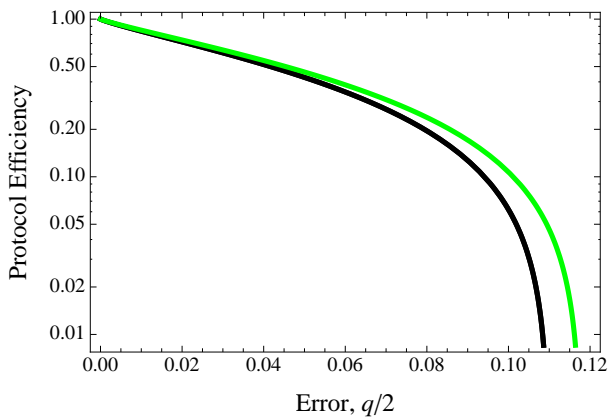} 
   \caption{Efficiency for Asymmetrical BB84, LM05 (both black) and SDC (green) against error rate $q/2$ for correlated channels specified by $e_1=e_2=e_m$  in the infinite key limit.}
   \label{fig8}
\end{figure} 
   
\noindent In both the figures we see LM05 outperforming BB84 though in the second only marginally. This can be understood as  due to going to infinite key limit where the two efficiencies should converge (binary entropic function for the bits to be discarded for both protocols in correlated channels is essentially given by the same form). This is illustrated in Figure (\ref{fig7}) for increasing qubits used and further highlighted in Figure (\ref{fig8}) where the black curve represents both the LM05 and BB84's key rate.  However what is interesting is the fact that for finite number of qubits used, LM05 exceeds BB84 in terms of a higher tolerance towards errors clearly exemplified in Figure (\ref{fig5}). This is almost unexpected as looking at the key rate formulas, one would imagine that the amount of bits to be discarded due to error correction and privacy amplification are essentially the same (relative to the key rates). However,
 a difference exists when considering the statistical fluctuation term. As LM05 allows for a larger number of bits to be salvaged in EM, thus a greater number of bits in CM (compared to BB84) may be used for error estimation in bounding Eve's information in the entropic relation. This allows for $\mu(n_e,k_e)<\mu(n,k)$ implying
\begin{eqnarray}
h_2\left (\dfrac{q}{2}+\mu(n,k)\right)>h_2\left (\dfrac{q}{2}+\mu(n_e,k_e)\right).
\end{eqnarray}
Hence, for the same error in a channel, BB84 discards more bits than LM05, explaining LM05's higher error tolerance. This would be essentially the same reasoning for LM05's performance over SDC's in figure \ref{fig4}.

SDC's better performance when compared to BB84's in both figures  and LM05's in figure \ref{fig5} is mainly due to the error correction term. In this particular model for correlated channels, the following inequality holds,
\begin{eqnarray}
h_2(q/2)\geq h_4(q/4),
\end{eqnarray}
where less bits are discarded for error correction in SDC and equality holds only for $q=0$. The infinite key regime of Figure (\ref{fig6}) sees identical performance for both LM05 and BB84 while SDC exceeds both.


\section{Conclusion}

\noindent Protocols making use of bidirectional quantum channels like the ones described in this work have the interesting feature of encodings being embedded in a unitary transformation as opposed to the preparation of a quantum state as well as the added advantage of how a decoding procedure should not see wastage due to bases mismatch unlike the conventional BB84. By conventional we refer to the BB84 in its original form where the choice of bases is equiprobable, resulting in only half of qubits transmitted could be used as a raw key. While asymmetric BB84 is proposed as a remedy and is expected to not perform any less than the two `deterministic' protocols in the infinite key regime, the scenario for finite keys can be different; thus the motivation for this work. 

Noting how the choice of relevant probability distribution for EM and CM is immediately analogous to the asymmetric BB84's choice for the preferred basis and non preferred basis respectively  in \cite{tom}, we extend the security proof based on purified versions of the SDC and LM05 in \cite{norm} to the use of smooth entropies for a finite key analysis. We provide for secure key rates in terms of efficiency for the protocols. 

In relatively small number of qubits resources used (order of $10^4$) as well as lower depolarization of channels, we observe an obvious advantage in LM05 due to having more bits for raw key purpose (derived from EM). This results from LM05's encoding/ decoding process which is independent of Bob's choice of bases for measurement processes; which is in fact the `deterministic' merit claim of the protocol. This advantage does however diminish in the region of asymptotically long keys when compared to BB84 (as well as the SDC).

The SDC is in some sense very similar to the asymmetric BB84 as bits for key purposes can only be salvaged when both Alice's and Bob's encoding and decoding processes respectively coincide (unlike LM05 where Bob's measurements in CM would be equally meaningful as a decoding measurement in the EM). The only real advantage SDC has over BB84 would be in the correlated channels case due to the error correction procedure. However, one must bear in mind that while efficiency for LM05 and BB84 is measured as key bits per-qubits transmitted, SDC's is measured as key bits per-entangled pairs used. Practically, this may not set to provide for a promising scenario as quantum memories become a necessity. 

We conclude saying that a finite key analysis would be enlightening also for two-way QKD in the framework of continuous variable \cite{Piran}. 

 
\section{Acknowledgement}  
The authors would like to thank N. Beaudry, S. Karumanchi and M. Lucamarini for fruitful discussions.
J. S. Shaari would also like to thank the University of Camerino for kind hospitality as well as the ICEO, IIUM for financial support.


\end{document}